\documentclass[12pt]{article}
\usepackage{epsf}

\newcommand{\EQ}{\begin{equation}}

\newcommand{\e}{{\rm e}}
\newcommand{\EN}{\end{equation}}

\newcommand{\bear}{\begin{eqnarray}}
\newcommand{\ear}{\end{eqnarray}}

\begin{document}

\topmargin 0pt
\oddsidemargin 5mm
\newcommand{\NP}[1]{Nucl.\ Phys.\ {\bf #1}}
\newcommand{\PL}[1]{Phys.\ Lett.\ {\bf #1}}
\newcommand{\NC}[1]{Nuovo Cimento {\bf #1}}
\newcommand{\CMP}[1]{Comm.\ Math.\ Phys.\ {\bf #1}}
\newcommand{\PR}[1]{Phys.\ Rev.\ {\bf #1}}
\newcommand{\PRL}[1]{Phys.\ Rev.\ Lett.\ {\bf #1}}
\newcommand{\MPL}[1]{Mod.\ Phys.\ Lett.\ {\bf #1}}
\newcommand{\JETP}[1]{Sov.\ Phys.\ JETP {\bf #1}}
\newcommand{\TMP}[1]{Teor.\ Mat.\ Fiz.\ {\bf #1}}
     
\renewcommand{\thefootnote}{\fnsymbol{footnote}}
     
\newpage
\setcounter{page}{0}
\begin{titlepage}     
IASSNS-HEP-00/31
\vspace{0.5cm}
\begin{center}
\large{  Exact and numerical results for the 
dimerized spin-orbital chain } \\
\vspace{1cm}
 {\large M.J. Martins$^{1,2}$  and B. Nienhuis$^{3}$} \\
\vspace{1cm}
\centerline{\em ${}^{1}$ School of Natural Sciences, Institute for Advanced Study}
\centerline{\em Olden Lane, Princeton, NJ 08540, USA}
\centerline{\em ${}^{2}$ Departamento de F\'isica, Universidade Federal de S\~ao Carlos}
\centerline{\em Caixa Postal 676, 13565-905, S\~ao Carlos, Brazil}
\centerline{\em ${}^{3}$ Instituut voor Theoretische Fysica, Universiteit van 
  Amsterdam}
\centerline{\em Valckenierstraat 65, 1018 XE, Amsterdam, The Netherlands}
\vspace{1.2cm}   
\end{center} 
\begin{abstract}
We establish exact results for the one-dimensional spin-orbital model for special
values of the four-spin interaction $V$ and dimerization parameter 
$\delta$. The
first exact result is at $\delta=1/2$ and $V=-2$. Because we find a very 
small but finite gap in this dimerized chain, this 
can serve as a very strong 
test case for numerical and approximate analytical techniques. 
The second result
is for the homogeneous chain with $V=-4$ and gives evidence that the system
has a spontaneously dimerized ground state. Numerical results indicate
that the interplay between dimerization and interaction could result in 
gapless phases in the regime $0\leq V<-2$.
\end{abstract}
\vspace{.2cm}
\centerline{PACS numbers: 05.50+q, 75.10.Jm }
\vspace{.2cm}
\centerline{March 2000}
\end{titlepage}

\renewcommand{\thefootnote}{\arabic{footnote}}

The renewed interest in spin-orbital models \cite{KK,CC} stems from
the belief that they are relevant to explain unusual magnetic properties of a wide range of
materials. Representative examples are quasi-one-dimensional TDAE-C$_{60}$
material \cite{AR}, the 
Na$_2$Ti$_2$SbO$_2$ and NaV$_2$O$_5$ compounds \cite{CO,RA},
and two-dimensional LINIO$_2$ samples \cite{LI}.
In this letter we study the dimerized spin-orbital chain with Hamiltonian
\begin{equation}
H=J\sum_{i=1}^{L} \sum_{\alpha=1,2}[1+\delta(-1)^{i+\alpha}] 
{\vec{S}}_{i}^{(\alpha)}. {\vec{S}}_{i+1}^{(\alpha)} + V\sum_{i=1}^{L}
(\vec{{S}}_{i}^{(1)}. {\vec{S}}_{i+1}^{(1)}) \times
({\vec{S}}_{i}^{(2)}. {\vec{S}}_{i+1}^{(2)}) 
\end{equation}
where ${\vec{S}}_{i}^{(\alpha)}$, $\alpha=1,2$ are two commuting sets of
$s=1/2$ operators acting on site $i$ of a lattice of size $L$. Notice
that the dimerization $\delta$ is completely staggered, which seems to be
the most interesting case at least in the context of spin
ladders \cite{DE,CA}. The the four-spin coupling $V$ in spin-orbital models 
originates in the standard superexchange interaction \cite{KK}. In spin ladders, it can be interpreted 
as a result of the Coulomb interaction between holes 
in a doped phase \cite{TE}. Here, unless stated otherwise, we assume an antiferromagnetic $J >0$ 
Heisenberg energy scale.  

Much of the theoretical efforts have been focused on the study of 
the homogeneous chain both numerically \cite{RA,NU1,NU2}
and analytically \cite{A1,A2}.  In general, the interesting features occur
in the strong coupling regime where perturbation theory does not work. In these cases it
is important to have access to exact results since
they play the role of relevant testing devices for
approximate non-perturbative methods.  One such 
example is the point $V/J=4$ in the homogeneous
chain, which is 
equivalent to the integrable SU(4) exchange spin
chain \cite{SU}. This fact has been not only useful to check numerical
analysis \cite{RA,NU1} but also relevant to trigger further
non-perturbative studies \cite{FU}. One of the purposes of this paper
is to point out that there exists two other points in which the model
(1) is still exactly solvable. Equally important, both results
are concerned with the strong coupling regime $|V| >> J$, namely
\begin{equation}
\mbox{(I)}~ \delta=1/2,~~V/J=-2~~~~\mbox{and}~~~~\mbox{(II)}~\delta=0,~~V/J=-4
\end{equation}

The exact solution enables us to derive analytical 
expressions for the ground state energy and the low-lying gap excitation. 
To our knowledge, the first case is a rare example where we can provide the exact value
of the energy gap for an interacting spin chain with {\em alternating} bond strength.
It should be emphasized that this result
is both at a feasible value
of the spin-Peierls dimerization and in a physically meaningful regime 
of the coupling $V$  \cite{TE}. 
The utility of these exact results are threefold. First of all, since
the lowest gap  excitation for case (I) is very small, it provides us
a relevant check of the reliability of numerical and approximate non-perturbative
methods to distinguish between gapped and gapless phases. 
Moreover, we present evidence
that the second case is an
example of a system with a spontaneous spin-Peierls effect. This lends 
support to the prediction by Nersesyan and Tsvelick \cite{TE} that
{\em negative}
four-spin interaction can induce
non-Haldane spin-liquid behaviour in spin ladders.  
The novelty here is that this
fact is established
at the very strong coupling regime $V/J=-4$, which is beyond the
reach of the approach of ref. \cite{TE}. Finally, 
together,  they motivate us
to study the gap behaviour for arbitrary values of dimerization
and interaction and to search for possible gapless phases.

The exact integrability is derived by 
identifying the lattice statistical
model whose row-to-row transfer matrix commutes with Hamiltonian (1) 
with the parameters
$\delta$ and $V/J$ given by (2). Such classical statistical system 
consists of two isotropic six vertex models coupled by the
total energy-energy interaction. The row-to-row transfer matrix $T(\lambda)$
of a vertex model
is usually written as the trace of an ordered product of Boltzmann weights,
\begin{equation}
T(\lambda)= Tr_{\cal{A}}[{\cal{L}}_{L,\cal{A}}^{(1)}(\lambda) 
{\cal{L}}_{L-1,\cal{A}}^{(2)}(\lambda) \cdots
{\cal{L}}_{2,\cal{A}}^{(1)}(\lambda) 
{\cal{L}}_{1,\cal{A}}^{(2)}(\lambda)]
\end{equation}
where 
${\cal{L}}_{j,\cal{A}}^{(\alpha)}(\lambda) $ denotes the local Boltzmann weights
with auxiliary space $\cal{A}$ and
quantum space $j=1,\cdots, L$, parametrized by the spectral
parameter $\lambda$. The auxiliary and quantum spaces correspond to the
horizontal and vertical degrees of freedom of the coupled six vertex models. In our
case there are 
four possible states per bond.

Essential to our approach is to notice that the Boltzmann 
weights can be conveniently written in terms
of two commuting
Temperly-Lieb operators $E_{j,{\cal A}}^{(\alpha)}= 2(
{\vec{S}}_{j}^{(\alpha)}. {\vec{S}}_{\cal{A}}^{(\alpha)}-I_{j,\cal{A}}/4)$, where
$I_{j,\cal{A}}$ is the identity operator.
More precisely, defining $
{\cal{L}}_{j,\cal{A}}^{(\alpha)}(\lambda) = P_{j,\cal{A}}
R_{j,\cal{A}}^{(\alpha)}(\lambda) $ with $ P_{j,\cal{A}}$ denoting the
exchange operator, we found that the $R$-matrix 
$R_{j,\cal{A}}^{(\alpha)}(\lambda) $ is given by
\begin{equation}
R_{j,\cal{A}}^{(\alpha)}(\lambda)= I_{j,\cal{A}} +\omega_1(\lambda) 
E_{j,\cal{A}}^{(\alpha)} +
\omega_2(\lambda) 
E_{j,\cal{A}}^{(1)} 
E_{j,\cal{A}}^{(2)} 
\end{equation}

The weight $\omega_2(\lambda)$ plays the role of the four-spin interaction
$V$ while a combination with $\omega_1(\lambda)$ is responsible for
the dimerization $\delta$. These weights belong to a more general class of
systems, that originally appeared in integrable coupled Potts models \cite{PER},
and were recently rediscovered in the context of Lorentz lattice gases 
and Fuss-Catalan algebras \cite{MB}. 
For the cases we are interested in, their explicit expressions are
\begin{eqnarray}
\omega_1(\lambda)= \left \{ \begin{array}{ll}
               \displaystyle{\frac{\e^{\lambda}-1}{2}} & \mbox{ for (I)} \\
               0 & \mbox{ for (II)}  \end{array} \right. ~~~{\rm and}~~~
\omega_2(\lambda)= \left \{ \begin{array}{ll}
     \displaystyle{\frac{\e^{\lambda}(\e^{\lambda}-1)}{3-\e^{\lambda}}} & \mbox{ for (I)} \\
     \displaystyle{\frac{\sinh(\lambda)}{\sinh(\gamma-\lambda)}} & \mbox{ for (II)} 
\end{array} \right.
\end{eqnarray}
where $\gamma=\ln(2+\sqrt{3})$.

As usual, the corresponding Hamiltonian is obtained as the first-order
expansion in $\lambda$ of the logarithm of $T(\lambda)$. By using
the weights (5) it is not difficult  to verify
that we indeed recover the Hamiltonian (1), 
up to irrelevant re-scaling constants, at the values given by (2).
To make further progress we have to explore other
properties of the transfer-matrix. In particular, we are interested
to establish the inversion relation \cite{SP,PE,KU}, since it in principle provides us
the means
to compute the ground state 
energy and excitation properties.
We will start by considering the most
involved model which is the case with non-null dimerization $\delta=1/2$. 
For this system, the inversion identity follows from  a combination
between the usual
unitary property, 
${\cal{L}}_{1,2}^{(\alpha)}(\lambda) 
{\cal{L}}_{2,1}^{(\alpha)}(-\lambda) ={\rm I}_{1,2}$,
and a less standard 
crossing relation for the weights
${\cal{L}}_{1,2}^{(\alpha)}(\lambda) $. These operators are not
independently crossing symmetric, instead they satisfy a novel
``mixed'' crossing property, which reads
\begin{equation} 
{\cal{L}}_{1,2}^{(\alpha)}(\lambda) = 
\frac{\omega_1(\lambda)}{\omega_1(\ln3-\lambda)} \stackrel{1}{M}
[{\cal{L}}_{1,2}^{(3-\alpha)}]^{t_2}(\ln3-\lambda)\stackrel{1}{M}  ,~~\alpha=1,2
\end{equation}
where  
$t_k$ denotes the transpose and $\stackrel{k}{M}$ is a $4 \times 4$ antidiagonal
matrix both acting on the space $k$.

Together with unitarity, this crossing relation allows us to derive
the following inversion identity,
\begin{equation}
T^{({\rm I})}(\lambda) T^{({\rm I})}(\lambda+\ln3)= \left [  
\frac{\omega_1(\ln3+\lambda)}{\omega_1(-\lambda)} \right ]^{L} \mbox{Id} +T^{+}(\lambda)
\end{equation}
where $T^{+}(\lambda)$  is a matrix whose elements for large L
are exponentially small, $O(\e^{-L})$, compared to the term
proportional to the identity Id. This means that,
in the thermodynamic limit, the last term in (7) vanishes, providing us
a much simpler functional equation for  all the transfer matrix eigenvalues.
In particular, the largest eigenvalue $\Lambda_{gs}^{({\rm I})}(\lambda)$ per site satisfies  
\begin{equation}
\Lambda_{gs}^{({\rm I})}(\lambda) \Lambda_{gs}^{({\rm I})}(\ln3+\lambda)=
\frac{\omega_1(\ln3+\lambda)}{\omega_1(-\lambda)} 
\end{equation}

With the help of unitarity, $\Lambda_{gs}^{({\rm I})}(\lambda) \Lambda_{gs}^{({\rm I})}(-\lambda)=1$,
it is possible to solve the functional equation (8) under plausible
analyticity assumption in the region
$0 \leq \lambda < \ln3$ where the Boltzmann weights are positive. The solution 
is given by
\begin{equation}
\Lambda_{gs}^{({\rm I})}(\lambda)= (\e^{\lambda}-1) \prod_{j=0}^{\infty}
\frac{(\e^{\lambda}3^{2j+1}-1)}{(\e^{\lambda}3^{2j}-1)}
\frac{(\e^{-\lambda}3^{2j+2}-1)}{(\e^{-\lambda}3^{2j+1}-1)}
\end{equation}

We now have the basic ingredients to derive the exact value for the ground
state energy per site $E_{gs}^{({\rm I})}/J$ 
of Hamiltonian (1) at $\delta=1/2$ and $V/J=-2$. This value is obtained
by computing $E_{gs}^{({\rm I})}/J=-\frac{d \ln \Lambda_{gs}^{({\rm I})}
(\lambda)}{d \lambda}|_{\lambda=0}+3/8$,
which reads
\begin{equation}
E_{gs}^{({\rm I})}/J= -\frac{5}{8} -12 \sum_{j=0}^{\infty} \frac{3^{2j}}{(3^{2j+1}-1)(3^{2j+2}-1)}
=-1.43312 6534
\end{equation}

 The inversion relation and trigonometric periodicity impose stringent
constraints to the form of the low-lying excitations. They should be described in terms
of meromorphic functions having two 
independent periods $2\ln(3)$ and $2 \pi$.
This observation alone 
enables us to calculate the dispersion relation \cite{KU} and from that
one obtains the exact value for the gap. Such gap corresponds to the energy 
necessary to create
an excitation with total spin $S^{z}=1$ in the spin-orbital model (I).
Here we omit further technicalities and
present only the final result. We found
that the triplet energy gap $\Delta^{({\rm I})}/J$ is given by
\begin{equation}
\frac{\Delta^{({\rm I})}}{J}= \prod_{j=1}^{\infty} \left [ \frac{(1-3^{-j/2})}{(1+3^{-j/2})}
\right ]^2 =0.00286 9614
\end{equation}

Interesting enough, the energy gap is very small and this  has 
an immediate application. It could be used to test
if a given
numerical or approximate non-perturbative method can really make a clear
distinction between a small gap and a real gapless phase. In general, this is
a difficult task, and we 
expect that our exact result will
be quite relevant to determine
suitability of multiprecision methods 
in spin ladders models. For instance, this should be important 
when one wants to predict the scaling
behaviour of the gap as a function of the dimerization \cite{TEN}.

We turn next to the second solvable point $\delta=0$ and $V/J=-4$.  In this case, 
$\omega_1(\lambda)$ is null and we are only left with the product
of two commuting isotropic six vertex models. From the point of view of the
classical statistical model, this means that we are in fact
dealing with an alternative representation of the sixteen-state
Potts model. By now several properties of the general $q$-state Potts
are fairly well understood. In particular, the ground state and the triplet
gap can be determined either by the inversion trick as above
\cite{KU} or by a direct mapping onto the $XXZ$ Heisenberg chain \cite{BA}.
Their exact expressions are
\begin{equation}
\frac{E_{gs}^{({\rm II})}}{J}= -\sqrt{3}\left [ 1+4\sum_{j=1}^{\infty} 
\frac{1}{1+(2+\sqrt{3})^{2j}} \right ] + \frac{1}{4} =-1.98444 4091
\end{equation}
and 
\begin{equation}
\frac{\Delta^{({\rm II})}}{J}= 2 \sqrt{3} \prod_{j=1}^{\infty} 
\left [ \frac{[1-(2+\sqrt{3})^{-j}]}{[1+(2+\sqrt{3})^{-j}]}
\right ]^2= 0.77960 4542
\end{equation}

Further interesting results can still be derived from the mapping
of the homogeneous spin-orbital model (1) at $V/J=-4$ onto the
the antiferromagnetic $XXZ$ chain with
anisotropy $J^{z}/J^{x}=2$ \cite{KU,BA}.  
First, it is possible to show that the
gap of the first excitation in 
the sector of total spin $S^{z}=0$ vanishes exponentially
as $L \rightarrow \infty$. 
The momentum of this excitation is $\pi$, which is compatible with the interpretation
that in the thermodynamic limit
the system has two spontaneously dimerized
ground states, in accordance  with the Lieb-Schultz-Mattis
theorem \cite{MSL,OS}. This result
supports the prediction by Nersesyan and Tsvelick \cite{TE} that 
a four-spin interaction may induce dimerized phases in spin ladder models.
The bosonization arguments of ref.\cite{TE} for
the weak coupling regime $|V|/J <<1$ together with our exact result
at the very strong coupling point $V/J=-4$  indicate that 
such dimerized phase should be robust for a rather large region of $V<0$.
Next, one can explain the numerical observations by 
Pati, Singh and Khomskii \cite{RA} that the model (II), {\em now} {\em for}
$J<0$ (point $B$ in figure 1 of ref. \cite{RA}), possesses an infinitely 
degenerated ground state. In fact, $J<0$ corresponds to the
{\em ferromagnetic} regime of the $XXZ$ chain, whose $T=0$ finite entropy
for $J^{z}/J^{x}=-2$ is exactly computed to be $\ln(2 +\sqrt{3})$. This value
is a rigorous confirmation of the 
lower bound proposed in ref. \cite{RA} for such residual entropy.

Considering these exact results, it is natural  to ask if a combined effect
of dimerization and $negative$ four-spin interaction could lead to a
gapless regime.
To investigate this problem we numerically
diagonalize the Hamiltonian (1), with periodic boundary conditions,
up to L=14 sites by using a Lanczos-type algorithm. In table 1 we 
exhibit our numerical results for the ground state energy and the
triplet energy gap for both cases (I) and (II). The extrapolations
towards the infinite volume limit 
was performed by using the Van den Broeck-Schwartz method
of convergence \cite{BW}. While the results
for the ground state are in good agreement with the exact values,
the gap estimates have a rather poor accuracy.
This emphasizes the importance of our exact results, 
since they clearly
show that  
one cannot trust 
the numerical gap estimates beyond two significant digits.
However, this does not prevent us to make a {\em qualitative}
comparison of the behaviour of the 
gap in the whole dimerization region $0 \leq \delta
\leq 1$ 
for various values of the interaction $V$.
This is illustrated in figure 1 for three values
of $V$ in the strong coupling regime. We 
observe that indeed for each value of $V/J$
the gap has a very small minimum for an appropriate value
of the parameter $\delta$.
Note that for $V/J=-2$  this minimum
occurs very near
the solvable point $\delta=1/2$ where the gap is small but still finite. 
This helps us to establish an 
upper bound for $V/J$, beyond which one probably should 
rule out strictly null mass gaps.
For $V/J<-2$ we observe, however, that this minimum decreases 
faster towards zero, suggesting the possibility of a  
gapless line on the plane $(\delta,
V/J)$ in the $0 \leq V/J<-2$ regime. It  would be
interesting to confirm the existence of this massless line
via 
more powerful numerical methods such as the
density matrix renormalization group \cite{WI}. In particular, this method  will 
allow us
to determine with good accuracy the critical value $V/J$ where the
gapless line starts. In practice, however, when a gap is as small as we
calculated, its effect would be invisible at 
even  low temperatures which may still be considerably higher than the value
of the gap.

In summary, we have pointed out the existence of two integrable points in
the spin-orbital model where relevant physical quantities such as
the ground state energy and the excitation gap can be evaluated exactly.
These results together with the solvable SU(4) symmetric point seem
to exhaust all possible Bethe ansatz integrable cases of Hamiltonian (1).
In addition, our numerical analysis indicate the possibility of a
gapless line on the plane $(\delta,V/J)$ for $0 \leq
 V/J<-2$. 
This then will add other example of coupled spin chains in which 
suitable combination of dimerization 
and interaction strength is capable to close the energy gap \cite{DE,CA}.
We also hope that our observations will motivate further numerical
and analytical investigation in related systems.

\section*{Acknowledgements}
We are grateful to A.L. Malvezzi and O.Techernyshyov for valuable discussions. 
This work was supported by Lampadia Foundation, by the Brazilian agencies CNPq and
Fapesp and by the Netherlands Foundation FOM.

\newpage
\underline{Table 1}: Finite size and extrapolated results for the 
ground state energy and the triplet gap for cases (I) and (II). For
comparison we also exhibit the exact results.

\begin{table}
\begin{center}
\begin{tabular}{|c|c|c|c|c|} \hline
L & $ E_{gs}^{({\rm I})}/J$ & $E_{gs}^{({\rm II})}/J$ & $\Delta^{({\rm I})}/J $ & $\Delta^{({\rm II})}/J$ \\ \hline\hline
4  & -1.625 & -2.25 & 2.267 949 & 4.  \\ \hline
6  & -1.510 455 & -2.083 333 & 1.529 642 & 2.876 894  \\ \hline
8  & -1.474 743 & -2.033 159 & 1.162 347 & 2.306 000  \\ \hline
10 & -1.459 071 & -2.012 015 & 0.939 894 & 1.957 187  \\ \hline
12 & -1.450 822 & -2.001 389 & 0.790 102 & 1.722 250  \\ \hline
14 & -1.445 954 & -1.995 437 & 0.682 160 & 1.553 984  \\ \hline
Extrap. & -1.433 1($\pm 1$) & -1.983 ($\pm 1$) & 0.011 ($\pm 3$)  & 0.744 ($\pm 2$) \\ \hline
Exact & -1.433 126  & -1.984 444  & 0.002 869  &  0.779 604 \\ \hline
\end{tabular}
\end{center}
\end{table}

\newpage
\begin{figure}[t]
\centerline{\epsfxsize=\textwidth\epsfbox{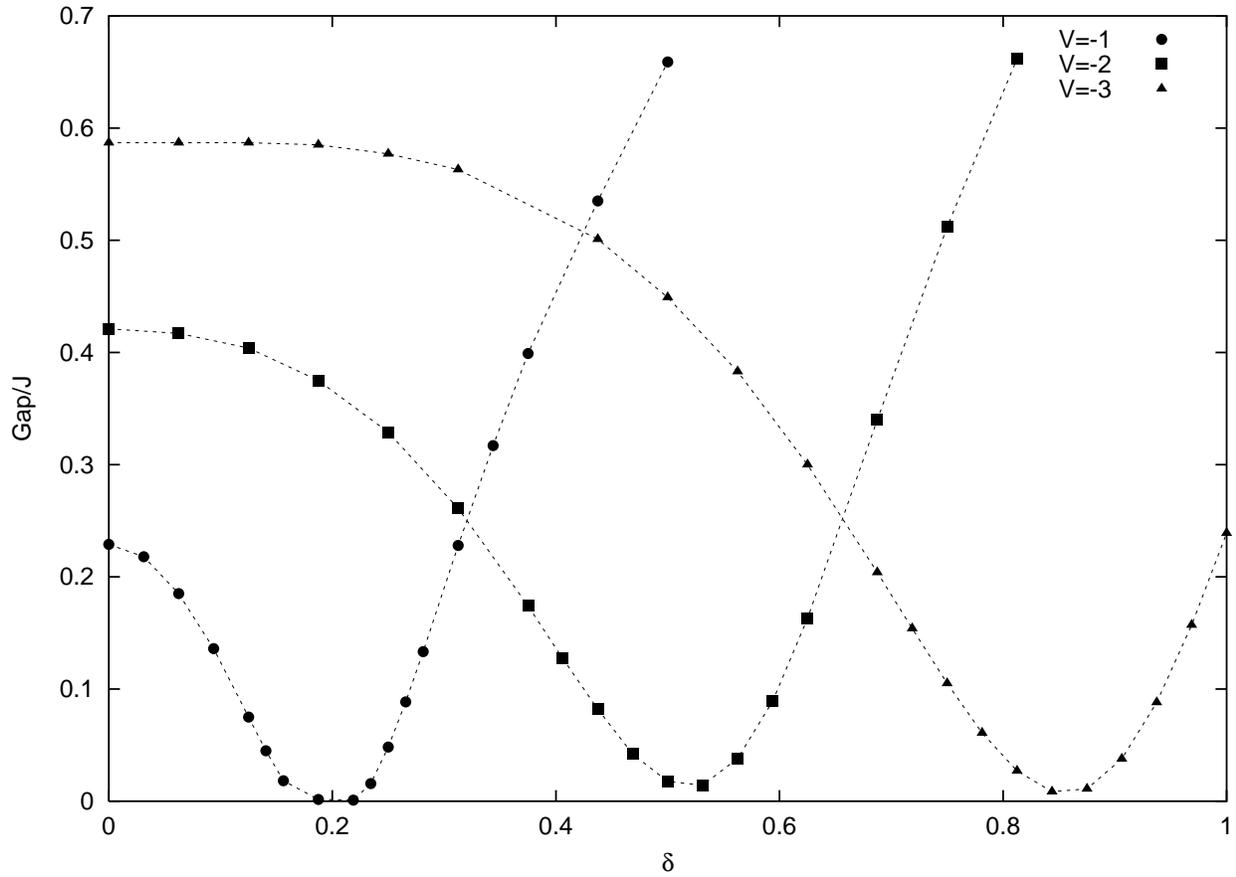}}
\vskip 5pt
\caption{ \protect{\footnotesize Extrapolated triplet energy gap $\Delta/J$
for $V/J=-1,-2,-3$. For $V/J \leq -4$ we no longer observed a local
minimum for the gap in the interval $0\leq \delta \leq 1$.
}}
\end{figure}

\end{document}